\documentclass[reprint,amsmath,amssymb,aps,pra]{revtex4-2}
\usepackage{listings}
\usepackage{graphicx}
\usepackage{dcolumn}
\usepackage{bm}
\usepackage{xspace}
\usepackage{todonotes}
\usepackage{color}
\newcommand{\RC}{Model Cards\xspace}

\begin{document}


\title{\RC for Quantum Technologies Reporting}

\author{M.J.~Everitt}
 \email{m.j.everitt@physics.org}
\affiliation{%
Quantalytics, Loughborough, UK.}%
 \homepage{https://quantalytics.co.uk}
 \author{Siyuan Ji}%
\affiliation{%
Quantum Systems Engineering Group, Loughborough University, Loughborough, LE11~3TU, UK.
}%
\begin{abstract}
There are a number of emerging quantum technologies that have the potential to be disruptive in application areas such as computation, communication and sensing.
In such a rapidly emerging field, there is a need for: transparency and accountability pertaining to devices, their performance, and limitations; the ability to assess new entities for integration into existing systems; sufficient information to undertake technology selection; share knowledge within and across institutions and discipline domains; manage risk and assure compliance with regulatory frameworks; drive innovation.
Here we propose \RC for documentation detailing use-cases and performance characteristics of entities for use in quantum technologies. 
Purpose of this document is therefore to stimulate discussion and begin to motivate the community to build a sufficient body of knowledge so that the most useful form of \RC can be developed and standardised. 
\end{abstract}
\maketitle

\section{Introduction}
There are a number of technologies emerging that make use of quantum phenomena to promise to deliver an advantage over their classical counterparts. 
There are the requirements that quantum technologies share with any emerging technology, such as stakeholders needing to understand the specific capabilities and limitations of each new contribution.
In addition to this there are a number of very specific challenges that are unique to quantum technologies.
Examples include: the verification of quantum algorithms for molecular simulation (unlike Shor's algorithm, there is no classically efficient check); the assurance of Quantum Key Distribution (QKD); guaranteeing performance within error budgets;  traceability and assurability of sensor data product outputs.
Many stakeholders with different levels of technical competence have a need for clarity on the intended use cases, performance characteristics, and evaluation conditions of products emerging in this space. 
This includes, but is not limited to, R\&D teams, systems engineers, and those in the supply chain from SME's through to policy makers. 
Currently, for quantum technologies, while a number of useful benchmarking metrics have been developed - such as various measures of quantum volume ~\cite{blume-kohout_volumetric_2020}, there are no standardised documentation procedures to communicate comparable benchmarks between quantum computers in a transparent and structured manner. 
We consider this document an unital proposal to stimulate discussion and note that a final form would need to be agreed by the community and informed by other approaches such as the ``methodology for power system domain experts to determine and describe their user requirements for automation systems'' (IEC~PAS~62559).

Quantum technologies share at least some assurance challenges with AI and machine learning. 
We propose a  solution to the  challenge of communicating device information that is based on the well-received example of a framework termed `Model Cards' from that community~\cite{RC}.
Our use of the word `Model' (paraphrasing the Oxford English dictionary) is a: \emph{A simplified or idealised description or conception of a particular system that is put forward as a basis for empirical understanding that may be used for calculations, predictions and evaluation}. 
We propose \RC as a mechanism for increasing transparency and assurability of quantum technologies. 
Acknowledging that different types of quantum technologies, such as computing, communication, and sensing, may have different needs and issues, we propose a generalised and extensible framework that could be adapted for different applications.
We note that extensibility is a fundamental requirement for these model cards. 
For example, there are current efforts to standardise the metrics and terminology for the characterisation of quantum technologies. 
Once they are available, a model card would very much benefit from ensuring compliance against the standard, as it is originally published, and as it evolves.

\section{Background \& motivation}
As noted in~\cite{RC} \emph{``Many mature industries have developed standardized methods of benchmarking various systems under different conditions.''} they cite the example of the electronic hardware industry providing data-sheets comprehensively characterising components’ performance metrics~\cite{gebru2021datasheets}. 
Another excellent example of how sets of documentation can be used to improve productivity was the International Technology Roadmap for Semiconductors (ITRS) and its successor, the International Roadmap for Devices and Systems (IRDS) which seeks to assess and document the current status and 15~year forecasted evolution of electronic devices and systems. 
By addressing topics such as application benchmarking, systems and architectures, heterogeneous integration, etc its ``intent is to provide a clear outline to simplify academic, manufacturing, supply, and research coordination regarding the development of electronic devices and systems''. 
It can be convincingly argued that such endeavours have been central to the past and future success of the semiconductor and related industries~\cite{10415924}.
While road-mapping activities do exist within quantum technologies, they do not have the same level of detail and activity as either the ITRS or the IDRS have benefited from~\cite{Acin_2018}. 
In our view, the reasons for this are at least in part due to the drivers for creating such road-maps being different, and as such the purpose they intend to, and can serve is also different. 

It is our view that by providing clarity on topics such as use cases, performance characteristics, and evaluation (including benchmarking) of quantum technologies \RC can support academic, manufacturing, supply, and research coordination stakeholders to accelerate the development and deployment of quantum devices and systems (as well as sub-components).
As such, the generation and sharing of \RC that we propose here may help bridge the gap between the high-level aspirations of the existing quantum technology road-maps and the immediate need to understand current capability and emulate successes in the ITRS and IDRS. 

As in the Model Card example of~\cite{RC}, the meaning and utility of quantum technologies will be somewhat subjective. For example:
\begin{itemize}
    \item End users could evaluate the `product' against its intended use cases and track its performance over time.
    \item Systems engineers could evaluate and model candidate components to assess their fitness for integration into wider systems.
    \item Policy makers and road-mappers could better track the development of the quantum technology ecosystem against technology-targets.
    \item Organisations can inform decisions or make more informed decisions about adopting a given quantum technology.
\end{itemize}
In all cases, the fundamental idea of providing a standardised document that details performance characteristics, intended use cases, and evaluation conditions applies directly to quantum technologies as much as it does to AI methods. 
Such transparency would enable stakeholders to better understand the capabilities and limitations of these technologies, build trust, and facilitate assurance processes.
We also note that, just as with AI methods, quantum computation (especially quantum machine learning) and sensing applications models should be evaluated for their ethical implications and, where relevant, inclusivity (e.g. in the emerging field of quantum machine learning) to ensure that the models do not contain bias or unrecognised inaccuracy.

The different use cases and nuances of quantum technologies mean that the cards proposed in~\cite{RC} cannot be used as-is and do need some substantive modification. They must acknowledge for instance: 
\begin{itemize}
    \item For performance evaluation, quantum computing will require benchmarks that reflect quantum-specific phenomena, such as quantum entanglement, superposition and decoherence, which do not have direct analogues in classical computing.
    \item In all quantum technologies, there are quantum-specific critical parameters that need to be captured in a \RC, such as coherence times, quantum bit error rate, or quantum volume.
    \item A clear articulation of assumptions made that may expose quantum vulnerabilities (e.g., quantum computer capabilities, physical channel security, sensitivity to atmospheric conditions, operating environment). All of which should be included in \RC, along with e.g. references to security proofs.
    \item As different realisations of a specific quantum technology modality may vary based on the choice of hardware, \RC should contain detailed information about its specific implementation.
    \item The effects of feedback and control on quantum systems are fundamentally different to those of classical systems. As such the \RC should pay particular attention to integration requirements and interface specifications.
    \item Where quantum algorithms may already be well-documented, the extent to which they are actually implemented on different hardware may vary significantly. This is especially true for simulators, such as those used for modelling molecular dynamics.
\end{itemize}

\section{\RC for Quantum Technologies}

Just as with AI Model Cards, Quantum technology \RC serve to disclose information about the quantum technology which we refer to as an \emph{'entity'} to emphasise the generality of devices and components to which this approach may be applied. The card should contain details of  what the entity asserts to achieve and what assumptions were made during its development. 
It should articulate the entity's expected behaviours for expected use cases. 
Inspired by~\cite{RC}, we propose a set of sections that quantum \RC should have which, just as for AI model cards, should include details that can inform all interested stakeholders for improved situational awareness and decision-making. 

As in~\cite{RC}, the proposed set of sections below are intended to as examples that should be tailored depending on use case. For this reason, we make a further recommendation beyond that of~\cite{RC} and assert that model cards be created using appropriate metadata conforming to FAIR Guiding Principles for scientific data management and stewardship~\cite{FAIR}. 
This would be of utility in e.g. supporting, potentially automated, road-mapping and trade-space exploration. 
Self-describing globally unique and persistent identifiers can be generated simply by using the hierarchy of the \RC and ideally would be agreed and then captured within the standardisation process and documentation. 
As far as possible the data captured within each card should also conform to the FAIR principles, such as by incorporating detailed provenance of the data (note that adoption of frameworks such as Units Markup language (UnitsML) to ensure that consistent and comparable recording of data in SI units would further help data mining from \RC).

\subsection{Entity Details}
This section includes the basic information needed to uniquely identify the entity. 
Uniqueness is important as, for example, it ensures that the entity and its components or systems of which it is a component are traceable through the supply chain. \\

\noindent\textbf{Name:} Name of the quantum entity.\\
\textbf{Version:} Version number or identifier.\\
\textbf{Type:} Type of quantum technology (e.g., computation, sensing).\\
\textbf{Purpose:} For example, \emph{``Simulator of a quantum spin liquid with configurable resonating valence bonds and full quantum state tomography''}, or \emph{``a magnetic flux sensor for non-destructive testing in of defects in aircraft composite structures''}, or \emph{``an implementation of BB84 for use in fibre-optical networks''}. To ensure conciseness but also provide clarity we recommend that this text conform to an ``18 Words Statement''~\cite{18W}.\\
\textbf{Developer:} May be company names or the names and affiliations of the developers.\\
\textbf{Release Date:} Date when the technology was released or documented.\\
\textbf{Supporting Documentation}: Examples include scientific publications, technical reports, drawings, patents, data-sets.\\
\textbf{Citation Details:} How should the model be cited?\\
\textbf{License:} License information if relevant.\\
\textbf{Feedback:} E.g., what is an email address that people may write to for further information or to report a deviation from the data with the Card?

\subsection{Intended Use}
By analogy with~\cite{RC} this section details the entity's intended purpose. It essentially comprises a set of use-case(s) that collectively forms the context for the entity and defines the `boundary' of the entity. It expands on the purpose in the entity Details and lists each application and interaction that is likely for the entity. \\ 
    
\noindent\textbf{Use Case \#}: Enumerated summary of each application scenario the entity intends to address. 
As with the purpose statement, this should be a high-level heading supported by additional details in the list below. 
To ensure conciseness but also provide clarity we recommend that each application listed here conform to an ``18 Words Statement''~\cite{18W}. 
Following this recommended approach, if the description of the specific application scenarios is considered too abstract or insufficient to convey the intended meaning, visual aids using well-defined graphical languages, e.g. SysML, should be included as supporting documents and listed in the entity Details section.
\begin{itemize}
\item \textbf{Taxonomy:} 
\begin{itemize}
    \item \emph{Classification}: Classification within the broader quantum technology landscape (e.g. gate-based quantum computing, single-photon detection).
    \item \emph{Categories}: Categories of use-cases the entity is designed for (e.g., optimisation, simulation, range-estimation, secure communication).
    \item \emph{Family}: Specification of the operational family or families the model's algorithm belongs to (e.g., variational algorithms, photonic systems).
\end{itemize}
\item \textbf{Users:} who is likely to interact with the entity, and for what purposes. This may include end-users or , e.g., those in a maintenance or repair capacity (and the level of knowledge that they might need to achieve those tasks).
\item \textbf{Classical Alternatives:} Provides a quantified (where possible) comparison against nearest alternatives.
\item \textbf{Quantum Alternatives:} Provides a quantified comparison against nearest non-classical alternatives.
\item \textbf{Out-of-scope Uses}: highlights technology that this entity might easily be confused with but whose functionality is in some way fundamentally different. 
\item \textbf{Limitations}: Contexts or conditions where the entity should not be used or is not reliable. Limitations accounted for here should be consistently reflected and elaborated in the Performance Metrics section.
\end{itemize}

\subsection{Factors}
Here~\cite{RC} propose factors that are relevant to machine learning models but not to quantum technologies. Unless machine learning models are used in the entity that is - in which case that structure should be included here. Examples where this might be the case are quantum machine learning, entities where machine learning was involved in the design process or applications such as sensing where machine learning forms part of the classical information processing system.


\subsection{Quantum Technology Specifications}
The purpose of this section is to detail the specific aspect(s) of quantum physics that is \emph{essential} to the \emph{fundamental operation} of the entity. 
Given the fragility of quantum states, it is important that each and every aspect of quantum operation is clearly identified. 
This is important as stakeholders may need such information to integrate the entity into a larger system or simply to ensure that the entity will function as intended in the environment into which it is to be deployed.
This should \emph{not} include any statements on performance as that will be covered later.

\subsubsection{System Architecture}
\noindent\textbf{Quantum Process/Algorithms: } Define the quantum processes or algorithms used to achieve the defined use cases. This includes quantum gates, quantum operations, and quantum algorithm's tailored to the system's capabilities.\\
\textbf{Circuit Design: }High-level description or schematics of the quantum circuit design (which might be a layer model for a sensor or QKD system).\\
\textbf{Physical Entanglement}: Description of how entanglement resources may be distributed across the physical (not logical) hardware. May include parasitical entanglements that are unwanted or need to be controlled.

\subsubsection{Hardware Specification}

\noindent\textbf{Carriers of Quantum Information:} An enumerated list (and frequency count) of each and every physical component that contains functionalised non-classical information (superconducting device realisation of a transom qubit, cavity resonator, photons). Circuit parameters for each component should be given.
For composite carries such as logical qubits, both the top-level `logical' and physical components should be listed separately. 
\\
\textbf{Local Quantum Coherence:} to each and every component listed above, a description of local quantum coherent superposition requirements and states that are involved.\\
\textbf{Non-local Quantum Coherence:} a description of all entanglement requirements and states that are involved (e.g. GHZ states between components of type 1 [homogeneous system entanglement], $W$ states between components \{1,4,5\} [heterogeneous system entanglement], two-mode squeezed vacuum state between channels \{3,7\} with $\zeta$ having configurable $\phi$ and $r$ within the region of $4$.).\\
\textbf{Entanglement Strategy:} Description of how entanglement is used in the model. This is to be distinguished from the above non-local coherence section as that was about what entanglement is present and this is how it is used.
\textbf{Measurement:} Information about each and every detector and the mode of detection (e.g., type, efficiency, dead-time, false-positive rates etc).\\
\textbf{Interconnects:} Medium used for quantum signal transmission between components and important parameters for estimating non-functional attributes of the entity such as efficiency, loss, decoherence and unwanted couplings (e.g., optical fibre between components \{1,3\}, free-space induction between components \{3,4\}).\\
\textbf{Control:} All control circuitry to each carrier and the mechanism by which they operate. It should be possible to estimate the decoherence effects of control circuitry, as well as the back action of the entity on the measurement device from this information.\\ 
\textbf{Control Feedback:} Making reference to measurement and control apparatus. Control feedback mechanisms should be detailed here.\\
\textbf{Operational Environment:} Required or optimal environmental conditions (e.g., temperature, magnetic field). This may include limitations on noise characteristics of any connected system or any other factor likely to affect the quantum coherence listed above.\\
\textbf{Non-Operational Environment:} a list of all identified environments that have the potential to damage the quantum information of the entity, and the mechanism by which they operate and the parameters by which they are characterised.\\
\textbf{Hardware Requirements:} Necessary hardware for operation, including any classical components not listed elsewhere, such as classical systems required for the control of the operational environment (e.g., vacuum systems, temperature control).

\subsubsection{Interface Specification}
\noindent \textbf{Data Type:} Description of the physical or logical data that is being transmitted, internally within the entity and externally with other systems.\\ 
\textbf{Data Handling:} Formats and protocols for how the interfacing data processing systems handle the transmitted data. \\ 
\textbf{Potential Issues:} May include back action, vulnerability to electrical interference, etc.\\ 
\textbf{Assumptions:} the artifact’s internal and external interface.\\ 
\textbf{Requirements:} e.g. on the level of circuit noise stability of power supplies. May make reference to compliance with standards.\\ 
\textbf{Software Requirements:} Software for data analysis.

\subsubsection{Other Approaches}
There are different frameworks for how a system specification can be captured and represented. The above strategy is considered a minimal set that covers the three key elements for any (quantum or classical) system architecture: namely, functional, structural, and interface definition. Other research, ~\cite{lowe}, has proposed a layer model for quantum computing. For instance, a system architecture that could be represented using three layers:
 \begin{itemize}
     \item \textbf{Abstract Layer}: Description of the high-level algorithmic structure and logical operations.
     \item \textbf{Physical Layer}: Details on the physical implementation of qubits and quantum gates, including error correction techniques.
     \item \textbf{Control Layer}: Information on how quantum operations are controlled and measured, including classical control systems.
 \end{itemize}
 
In formulating layer models, principles such as those used by OSI (fewest possible layers, minimisation of interactions, maximisation of differences, etc.) will be of utility. The choice of the factors such as taxonomies, complexity class, quantum information class, error mitigation's and competing architectures all complicate this picture. We propose that the use of \RC may actually help in  the reverse process of helping the community build taxonomies and layer models that can be standardised.

\subsection{Errors}
The purpose of this section is for users to estimate if the overall reliability of the entity or system into which it is incorporated lies within their error budget. As such it is of relevance to product engineers, systems engineers, product managers, business analysts, and policy makers. Efforts should be made to make the information in this section both accurate and accessible.

\noindent\textbf{Source Identification:} Potential sources of error. Quantum systems may have non-classical failure modes. For example, superconductors with micro-fractures might experience Josephson effects that would modify the critical parameters of a superconducting quantum interference device to produce unexpected behaviours \\
\textbf{Classification systems:} In addition to the source of error, the form it takes should be clearly articulated as being quantum or classical in nature and the effect on given phenomena such as entanglement resources also be detailed. This might be effects such as the false-positive rate of a photon counter or the overestimate of a given entanglement measure. \\
\textbf{Impact Estimates:} Quantitative or qualitative impact of each error source. \\
\textbf{Mitigation Strategies:} Techniques to minimize errors, implemented or recommended.\\
\textbf{Residual Errors:} Expected impact on performance. \\

Information covered by the above points is inherently connected through casual chains, e.g., the occurrence of an identified error leading to a functional failure of the entity. To help visualise these casual relationships in a more structured and detailed manner, we recommended that a \emph{two-levels} Failure Mode and Effects Analysis (FMEA) is conducted to ensure failure modes associated with each and every quantum hardware specified in Section D-2 is identified and their effects at both the \emph{entity level} and the \emph{wider system level} (in which the entity is interacting with other classical systems specified in Section D-2). The FMEA should be included as a supporting document, referenced in Section A.

\subsection{Performance Metrics}
This section contains a list of model performance metrics and the reason they are selected over other possible measures of model performance.
It is important that the methodology for taking measurements and estimating the metrics is detailed and any uncertainty in the metric is also quantified. For example, this may include standard deviation, variance, confidence intervals, and details of how these values are arrived at also be included.

Metrics that may be relevant include, but are not limited to: benchmark tests,
comparison against classical equivalents, error rates, sensitivity, accuracy, precision, sampling frequency, quantum advantage, quantum bit error rate (QBER), secret key rate, maximum secure distance, stability, critical parameters, decoherence, dephasing and loss (such as T1 and T2).

Following~\cite{TPM} the metrics of interest here should be those that (a) relate to high-priority requirements (b) pertain to requirements that have high-risk priority number (RPN) (c) where there is a need to assess against performance targets (e.g. for control or assessment of technology development). These metrics may be displayed in any suitable form including graphically.

As noted in~\cite{TPM} the quantitative evaluation of performance can be expensive so it will usually not be practical to capture all possible performance metrics. 
For the purpose of \RC, those most important and relevant metrics for each listed use case should be provided.
We advise that metrics are reported using the following structure:
\begin{itemize}
    \item Name of metric
    \item Purpose: How it should be used.
    \item Use-case: Use-case \#s in which the metric is applied to.
    \item Definition: the metric (formulaic or algorithmic), its units, associated models, and the measured quantities used to calculate it.
    \item Risk level: as measured through RPN or some other accepted metric.
    \item Measurement: what was actually been measured and how it has been measured.
    \item Statistics: calculated metrics and associated sufficient details to establish confidence that the metrics presented establish their value.
    \item Benchmarks: against which the performance of the entity should be compared.
    \item Fundamental limit: The theoretical best possible performance against which performance should be measured.
\end{itemize}
Those metrics that may be of interest but are not captured may be listed at the end of the section.

\subsection{Ethical Considerations}
\noindent\textbf{Impact Assessment:} Potential ethical impacts such as a description of how the model handles data privacy and security.\\
\textbf{Mitigation Strategies:} Strategies to mitigate negative ethical impacts.

\subsection{Evaluation Criteria}
\noindent \textbf{Model Validation}: Steps taken to validate the model's performance, including cross-validation techniques or comparison with theoretical predictions.
\\ 
\textbf{Hardware Verification}: Description of the process to ensure that the quantum hardware used meets the necessary specifications and performance benchmarks.
\\
\textbf{Algorithmic Correctness}: Methods used to verify the correctness of the quantum algorithm, including simulation results or analytical proofs.
\\
\textbf{Operational Stability}: Evidence of the model's stability over time, including measures taken to mitigate decoherence and other quantum-specific issues.
\\
\textbf{Reproducibility}: Information on how to reproduce the model's results, including access to data sets, code, and detailed experimental procedures.

\subsection{Assurability}
Assurability in this context means providing stakeholders with confidence that the system meets the claimed mode of operation and performance standards in the other section of the \RC under the various conditions the entity is claimed to operate under. Assurance of a quantum system goes beyond evaluation as detailed in the previous section and is mandatory when the quantum entity is incorporated into a mission-, safety-, or security- critical system. Evidence required for assurance could differ between different types of quantum technologies. For instance, security measures are vital for the assurance of QKD. A non-exhaustive list of possible evidence that could be included in \RC is provided below. Alternatively, the authors of \RC could also include an Assurance Case~\cite{wei2019model} as a supporting document.

\noindent\textbf{Certifications}: List any certifications the entity has received that attest to its security or reliability.
\\
\textbf{Standards Compliance}: Describe compliance with international or industry-specific standards for quantum communication and security.
\\
\noindent \textbf{Audit Reports}: Summarize findings from independent audits of the system, focusing on security and performance evaluations.
\\
\textbf{Evaluation Partners}: List organizations or entities that have independently evaluated the system.
\\
\noindent \textbf{Physical Security}: Describe measures in place to ensure the physical security of the hardware and infrastructure.
\\ 
\textbf{Cybersecurity}: Outline cybersecurity practices and protocols to protect the system against digital threats.
\\
\noindent \textbf{Fail-Safe Mechanisms}: Detail any mechanisms or features designed to secure the system in the event of a failure or compromise.
 \\ 
 \textbf{Recovery Protocols}: Describe protocols for recovering from security breaches or system failures, e.g., key revocation and regeneration procedures in the case of an intercept-resend attack~\cite{PhysRevLett.98.030503}.

\noindent \textbf{Training Programs}: Information on training programs  available for users to ensure they can operate the system securely and effectively.
\\ 
\textbf{Support Resources}: Details on support resources, including technical support and documentation available to users.

\subsection{Supplementary Materials}

\noindent\textbf{References:} Key references and further reading.
\\ 
\textbf{Supporting Documents:} As listed in Section A, and referenced in the other sections.

\section{Concluding remarks}
 
We have proposed extending the idea of Model Cards for machine learning models to a general \RC for emerging quantum technology reporting.
\RC include information about the operation and performance of the quantum entities. 
The next is to this reporting structure based on feedback obtained from a range of stakeholders including technology regulators and end users.
One strong omission in our current proposal is that of data provenance and how that would be best achieved in a framework such as this. 
The framework presented here is intended to be general enough to inform technology selection, assess risk and inform policy. 
The usefulness and accuracy of this approach will depend on those engaging with it to act with transparency and include sufficient data on assurability, verification, and validation.

\begin{acknowledgments}
We gratefully acknowledge funding from Innovate UK 10102791 Scalable Quantum Network Technology. 
We also thank Cathy White, Douglas Paul, John Devaney and Derwen Hinds for useful suggestions and interesting and informative discussions.
\end{acknowledgments}

\section{Author Contributions}
\noindent Mark Everitt: Original idea, Conceptualization, Investigation, Writing - Original Draft. \\
\noindent Siyuan Ji: Conceptualization, Writing - Review \& Editing.

\bibliography{refs}

\begin{thebibliography}{10}%
\makeatletter
\providecommand \@ifxundefined [1]{%
 \@ifx{#1\undefined}
}%
\providecommand \@ifnum [1]{%
 \ifnum #1\expandafter \@firstoftwo
 \else \expandafter \@secondoftwo
 \fi
}%
\providecommand \@ifx [1]{%
 \ifx #1\expandafter \@firstoftwo
 \else \expandafter \@secondoftwo
 \fi
}%
\providecommand \natexlab [1]{#1}%
\providecommand \enquote  [1]{``#1''}%
\providecommand \bibnamefont  [1]{#1}%
\providecommand \bibfnamefont [1]{#1}%
\providecommand \citenamefont [1]{#1}%
\providecommand \href@noop [0]{\@secondoftwo}%
\providecommand \href [0]{\begingroup \@sanitize@url \@href}%
\providecommand \@href[1]{\@@startlink{#1}\@@href}%
\providecommand \@@href[1]{\endgroup#1\@@endlink}%
\providecommand \@sanitize@url [0]{\catcode `\\12\catcode `\$12\catcode
  `\&12\catcode `\#12\catcode `\^12\catcode `\_12\catcode `\%12\relax}%
\providecommand \@@startlink[1]{}%
\providecommand \@@endlink[0]{}%
\providecommand \url  [0]{\begingroup\@sanitize@url \@url }%
\providecommand \@url [1]{\endgroup\@href {#1}{\urlprefix }}%
\providecommand \urlprefix  [0]{URL }%
\providecommand \Eprint [0]{\href }%
\providecommand \doibase [0]{https://doi.org/}%
\providecommand \selectlanguage [0]{\@gobble}%
\providecommand \bibinfo  [0]{\@secondoftwo}%
\providecommand \bibfield  [0]{\@secondoftwo}%
\providecommand \translation [1]{[#1]}%
\providecommand \BibitemOpen [0]{}%
\providecommand \bibitemStop [0]{}%
\providecommand \bibitemNoStop [0]{.\EOS\space}%
\providecommand \EOS [0]{\spacefactor3000\relax}%
\providecommand \BibitemShut  [1]{\csname bibitem#1\endcsname}%
\let\auto@bib@innerbib\@empty
\bibitem [{\citenamefont {Blume-Kohout}\ and\ \citenamefont
  {Young}(2020)}]{blume-kohout_volumetric_2020}%
  \BibitemOpen
  \bibfield  {author} {\bibinfo {author} {\bibfnamefont {R.}~\bibnamefont
  {Blume-Kohout}}\ and\ \bibinfo {author} {\bibfnamefont {K.~C.}\ \bibnamefont
  {Young}},\ }\bibfield  {title} {\bibinfo {title} {A volumetric framework for
  quantum computer benchmarks},\ }\href
  {https://doi.org/10.22331/q-2020-11-15-362} {\bibfield  {journal} {\bibinfo
  {journal} {Quantum}\ }\textbf {\bibinfo {volume} {4}},\ \bibinfo {pages}
  {362} (\bibinfo {year} {2020})}\BibitemShut {NoStop}%
\bibitem [{\citenamefont {Mitchell}\ \emph {et~al.}(2019)\citenamefont
  {Mitchell}, \citenamefont {Wu}, \citenamefont {Zaldivar}, \citenamefont
  {Barnes}, \citenamefont {Vasserman}, \citenamefont {Hutchinson},
  \citenamefont {Spitzer}, \citenamefont {Raji},\ and\ \citenamefont
  {Gebru}}]{RC}%
  \BibitemOpen
  \bibfield  {author} {\bibinfo {author} {\bibfnamefont {M.}~\bibnamefont
  {Mitchell}}, \bibinfo {author} {\bibfnamefont {S.}~\bibnamefont {Wu}},
  \bibinfo {author} {\bibfnamefont {A.}~\bibnamefont {Zaldivar}}, \bibinfo
  {author} {\bibfnamefont {P.}~\bibnamefont {Barnes}}, \bibinfo {author}
  {\bibfnamefont {L.}~\bibnamefont {Vasserman}}, \bibinfo {author}
  {\bibfnamefont {B.}~\bibnamefont {Hutchinson}}, \bibinfo {author}
  {\bibfnamefont {E.}~\bibnamefont {Spitzer}}, \bibinfo {author} {\bibfnamefont
  {I.~D.}\ \bibnamefont {Raji}},\ and\ \bibinfo {author} {\bibfnamefont
  {T.}~\bibnamefont {Gebru}},\ }\bibfield  {title} {\bibinfo {title} {Model
  cards for model reporting},\ }in\ \href
  {https://doi.org/10.1145/3287560.3287596} {\emph {\bibinfo {booktitle}
  {Proceedings of the Conference on Fairness, Accountability, and
  Transparency}}},\ \bibinfo {series and number} {FAT* '19}\ (\bibinfo
  {publisher} {Association for Computing Machinery},\ \bibinfo {address} {New
  York, NY, USA},\ \bibinfo {year} {2019})\ p.\ \bibinfo {pages}
  {220–229}\BibitemShut {NoStop}%
\bibitem [{\citenamefont {Gebru}\ \emph {et~al.}(2021)\citenamefont {Gebru},
  \citenamefont {Morgenstern}, \citenamefont {Vecchione}, \citenamefont
  {Vaughan}, \citenamefont {Wallach}, \citenamefont {au2},\ and\ \citenamefont
  {Crawford}}]{gebru2021datasheets}%
  \BibitemOpen
  \bibfield  {author} {\bibinfo {author} {\bibfnamefont {T.}~\bibnamefont
  {Gebru}}, \bibinfo {author} {\bibfnamefont {J.}~\bibnamefont {Morgenstern}},
  \bibinfo {author} {\bibfnamefont {B.}~\bibnamefont {Vecchione}}, \bibinfo
  {author} {\bibfnamefont {J.~W.}\ \bibnamefont {Vaughan}}, \bibinfo {author}
  {\bibfnamefont {H.}~\bibnamefont {Wallach}}, \bibinfo {author} {\bibfnamefont
  {H.~D.~I.}\ \bibnamefont {au2}},\ and\ \bibinfo {author} {\bibfnamefont
  {K.}~\bibnamefont {Crawford}},\ }\href@noop {} {\bibinfo {title} {Datasheets
  for datasets}} (\bibinfo {year} {2021}),\ \Eprint
  {https://arxiv.org/abs/1803.09010} {arXiv:1803.09010 [cs.DB]} \BibitemShut
  {NoStop}%
\bibitem [{\citenamefont {Gargini}(2023)}]{10415924}%
  \BibitemOpen
  \bibfield  {author} {\bibinfo {author} {\bibfnamefont {P.~A.}\ \bibnamefont
  {Gargini}},\ }\bibfield  {title} {\bibinfo {title} {Overcoming semiconductor
  and electronics crises with irds: Planning for the future},\ }\href
  {https://doi.org/10.1109/MED.2023.3340123} {\bibfield  {journal} {\bibinfo
  {journal} {IEEE Electron Devices Magazine}\ }\textbf {\bibinfo {volume}
  {1}},\ \bibinfo {pages} {32} (\bibinfo {year} {2023})}\BibitemShut {NoStop}%
\bibitem [{\citenamefont {Wilkinson}\ \emph {et~al.}(2016)\citenamefont
  {Wilkinson}, \citenamefont {Dumontier}, \citenamefont {Aalbersberg},
  \citenamefont {Appleton}, \citenamefont {Axton}, \citenamefont {Baak},
  \citenamefont {Blomberg}, \citenamefont {Boiten}, \citenamefont
  {da~Silva~Santos}, \citenamefont {Bourne},\ and\ \citenamefont
  {et~al.}}]{FAIR}%
  \BibitemOpen
  \bibfield  {author} {\bibinfo {author} {\bibfnamefont {M.~D.}\ \bibnamefont
  {Wilkinson}}, \bibinfo {author} {\bibfnamefont {M.}~\bibnamefont
  {Dumontier}}, \bibinfo {author} {\bibfnamefont {I.~J.}\ \bibnamefont
  {Aalbersberg}}, \bibinfo {author} {\bibfnamefont {G.}~\bibnamefont
  {Appleton}}, \bibinfo {author} {\bibfnamefont {M.}~\bibnamefont {Axton}},
  \bibinfo {author} {\bibfnamefont {A.}~\bibnamefont {Baak}}, \bibinfo {author}
  {\bibfnamefont {N.}~\bibnamefont {Blomberg}}, \bibinfo {author}
  {\bibfnamefont {J.-W.}\ \bibnamefont {Boiten}}, \bibinfo {author}
  {\bibfnamefont {L.~B.}\ \bibnamefont {da~Silva~Santos}}, \bibinfo {author}
  {\bibfnamefont {P.~E.}\ \bibnamefont {Bourne}},\ and\ \bibinfo {author}
  {\bibnamefont {et~al.}},\ }\bibfield  {title} {\bibinfo {title} {The fair
  guiding principles for scientific data management and stewardship},\ }\href
  {https://www.nature.com/articles/sdata201618} {\bibfield  {journal} {\bibinfo
   {journal} {Nature News}\ } (\bibinfo {year} {2016})}\BibitemShut {NoStop}%
\bibitem [{\citenamefont {Burge}(2015)}]{18W}%
  \BibitemOpen
  \bibfield  {author} {\bibinfo {author} {\bibfnamefont {S.}~\bibnamefont
  {Burge}},\ }\href
  {https://www.burgehugheswalsh.co.uk/Uploaded/1/Documents/18-Word-Statement-Tool-v1.pdf}
  {\bibinfo {title} {18 words statement (the systems thinking tool box)}}
  (\bibinfo {year} {2015})\BibitemShut {NoStop}%
\bibitem [{\citenamefont {Lowe}(2022)}]{lowe}%
  \BibitemOpen
  \bibfield  {author} {\bibinfo {author} {\bibfnamefont {D.}~\bibnamefont
  {Lowe}},\ }\emph {\bibinfo {title} {Engineering Design and Quantum Computing
  (subbmtted version)}},\ \href@noop {} {Ph.D. thesis},\ \bibinfo  {school}
  {Loughborough University} (\bibinfo {year} {2022})\BibitemShut {NoStop}%
\bibitem [{\citenamefont {Oakes}\ \emph {et~al.}(2006)\citenamefont {Oakes},
  \citenamefont {Botta},\ and\ \citenamefont {Bahill}}]{TPM}%
  \BibitemOpen
  \bibfield  {author} {\bibinfo {author} {\bibfnamefont {J.}~\bibnamefont
  {Oakes}}, \bibinfo {author} {\bibfnamefont {R.}~\bibnamefont {Botta}},\ and\
  \bibinfo {author} {\bibfnamefont {T.}~\bibnamefont {Bahill}},\ }\bibfield
  {title} {\bibinfo {title} {11.1.1 technical performance measures},\ }\href
  {https://doi.org/https://doi.org/10.1002/j.2334-5837.2006.tb02826.x}
  {\bibfield  {journal} {\bibinfo  {journal} {INCOSE International Symposium}\
  }\textbf {\bibinfo {volume} {16}},\ \bibinfo {pages} {1466} (\bibinfo {year}
  {2006})},\ \Eprint
  {https://arxiv.org/abs/https://incose.onlinelibrary.wiley.com/doi/pdf/10.1002/j.2334-5837.2006.tb02826.x}
  {https://incose.onlinelibrary.wiley.com/doi/pdf/10.1002/j.2334-5837.2006.tb02826.x}
  \BibitemShut {NoStop}%
\bibitem [{\citenamefont {Wei}\ \emph {et~al.}(2019)\citenamefont {Wei},
  \citenamefont {Kelly}, \citenamefont {Dai}, \citenamefont {Zhao},\ and\
  \citenamefont {Hawkins}}]{wei2019model}%
  \BibitemOpen
  \bibfield  {author} {\bibinfo {author} {\bibfnamefont {R.}~\bibnamefont
  {Wei}}, \bibinfo {author} {\bibfnamefont {T.~P.}\ \bibnamefont {Kelly}},
  \bibinfo {author} {\bibfnamefont {X.}~\bibnamefont {Dai}}, \bibinfo {author}
  {\bibfnamefont {S.}~\bibnamefont {Zhao}},\ and\ \bibinfo {author}
  {\bibfnamefont {R.}~\bibnamefont {Hawkins}},\ }\bibfield  {title} {\bibinfo
  {title} {Model based system assurance using the structured assurance case
  metamodel},\ }\href@noop {} {\bibfield  {journal} {\bibinfo  {journal}
  {Journal of Systems and Software}\ }\textbf {\bibinfo {volume} {154}},\
  \bibinfo {pages} {211} (\bibinfo {year} {2019})}\BibitemShut {NoStop}%
\bibitem [{\citenamefont {Lodewyck}\ \emph {et~al.}(2007)\citenamefont
  {Lodewyck}, \citenamefont {Debuisschert}, \citenamefont
  {Garc\'{\i}a-Patr\'on}, \citenamefont {Tualle-Brouri}, \citenamefont {Cerf},\
  and\ \citenamefont {Grangier}}]{PhysRevLett.98.030503}%
  \BibitemOpen
  \bibfield  {author} {\bibinfo {author} {\bibfnamefont {J.}~\bibnamefont
  {Lodewyck}}, \bibinfo {author} {\bibfnamefont {T.}~\bibnamefont
  {Debuisschert}}, \bibinfo {author} {\bibfnamefont {R.}~\bibnamefont
  {Garc\'{\i}a-Patr\'on}}, \bibinfo {author} {\bibfnamefont {R.}~\bibnamefont
  {Tualle-Brouri}}, \bibinfo {author} {\bibfnamefont {N.~J.}\ \bibnamefont
  {Cerf}},\ and\ \bibinfo {author} {\bibfnamefont {P.}~\bibnamefont
  {Grangier}},\ }\bibfield  {title} {\bibinfo {title} {Experimental
  implementation of non-gaussian attacks on a continuous-variable
  quantum-key-distribution system},\ }\href
  {https://doi.org/10.1103/PhysRevLett.98.030503} {\bibfield  {journal}
  {\bibinfo  {journal} {Phys. Rev. Lett.}\ }\textbf {\bibinfo {volume} {98}},\
  \bibinfo {pages} {030503} (\bibinfo {year} {2007})}\BibitemShut {NoStop}%
\end{thebibliography}%

\end{document}